\documentclass[sigconf,screen=true,anonymous=false,bookmarks=false,9pt]{acmart}
\usepackage{siunitx}

\usepackage[htt]{hyphenat}

\usepackage{titlesec}
\titlespacing\section{0pt}{4pt plus 1pt minus 1pt}{0pt plus 1pt minus 1pt}
\titlespacing\subsection{0pt}{4pt plus 1pt minus 1pt}{1pt plus 1pt minus 1pt}
\titlespacing\subsubsection{0pt}{4pt plus 1pt minus 1pt}{2pt plus 1pt minus 1pt}
\makeatletter
\@ifundefined{ACM@origsection}{}{\let\ACM@origsection\section}
\@ifundefined{ACM@origsubsection}{}{\let\ACM@origsubsection\subsection}
\@ifundefined{ACM@origsubsubsection}{}{\let\ACM@origsubsubsection\subsubsection}
\@ifundefined{ACM@origparagraph}{}{\let\ACM@origparagraph\paragraph}
\makeatother

\iftrue
\fi

\usepackage{pifont}
\newcommand{\cmark}{\ding{51}}
\newcommand{\xmark}{\ding{55}}

\usepackage{lipsum}
\usepackage{graphicx}
\usepackage{amsmath}

\usepackage{footnote}
\usepackage{mathtools}
\usepackage{mathrsfs}     
\usepackage{comment}
\usepackage[subrefformat=parens,labelformat=parens]{subfig}
\usepackage{bm}
\usepackage{multirow}
\usepackage{threeparttable,booktabs}
\usepackage{blkarray}
\usepackage{tikz}
\usetikzlibrary{positioning,calc,fit,decorations.pathmorphing,shapes.geometric,shapes.gates.logic.US,calc}
\usepackage{balance}
\usepackage{courier}                                       
\usepackage{cleveref}                                      
\usepackage[mathcal]{eucal}
\usepackage[]{algpseudocode}                               
\algrenewcommand\textproc{\texttt}
\makeatletter\let\float@addtolists\relax\makeatother
\usepackage{algorithm}
\usepackage{filecontents}                                  
\usepackage{pgfplots}
\usepackage{pgfplotstable}
\usepackage{makecell}
\pgfplotsset{compat=newest}
\usepackage[figuresright]{rotating}

\usepackage[inline]{enumitem}                              
\setlist{leftmargin=5.08mm}

\theoremstyle{plain}

\theoremstyle{definition}

\algrenewcommand\textproc{\texttt}

\usepackage[skip=1pt]{caption}            
\definecolor{CUHKorange}{RGB}{244,106,18} 
\definecolor{CUHKblue}{RGB}{0,111,190}    
\definecolor{CUHKgreen}{RGB}{0,127,128}   
\definecolor{CUHKred}{RGB}{228,46,36}     
\definecolor{CUHKyellow}{RGB}{198,148,34} 
\definecolor{CUHKdark}{RGB}{114,44,114}   
\definecolor{CUHKmiddle}{RGB}{144,44,144} 

\usepackage{tcolorbox}
\tcbuselibrary{skins,breakable}
    {\endtcolorbox}

\usepackage{tikz}
\usepackage{pgfplots}
\pgfplotsset{compat=1.17}
\usepgfplotslibrary{groupplots} 
\usepackage{pgfplotstable}

\graphicspath{{./figs/}}

\copyrightyear{2026}
\acmYear{2026}
\setcopyright{cc}
\setcctype{by}
\acmConference[ICCAD '26]{IEEE/ACM International Conference on Computer-Aided Design}{November 08--12, 2026}{San Jose, CA, USA}
\acmBooktitle{IEEE/ACM International Conference on Computer-Aided Design (ICCAD '26), November 08--12, 2026, San Jose, CA, USA}
\acmDOI{10.1145/3831252.3834022}
\acmISBN{979-8-4007-2873-0/2026/11}

\begin{document}
\date{}

\title{Tri-Stack-PIM: A Heterogeneous Multi-Die 3D-Integrated Processing-in-Memory Architecture for Efficient LLM Inference}

\author{Shuao Jia}
\orcid{0009-0005-8753-4068}
\affiliation{%
    \institution{Beijing University of Posts and Telecommunications}
    \city{Beijing}
    \country{China}
}
\email{pdxqjiashuao@bupt.edu.cn}

\author{Zichao Ling}
\orcid{0009-0002-0942-6991}
\affiliation{%
    \institution{Beijing University of Posts and Telecommunications}
    \city{Beijing}
    \country{China}
}
\email{lingzichao@bupt.edu.cn}

\author{Chen Bai}
\orcid{0000-0002-1742-0090}
\affiliation{%
    \institution{Fudan University}
    \city{Shanghai}
    \country{China}
}
\email{cbai@fudan.edu.cn}

\author{Kang Zhao}
\orcid{0000-0003-0502-8523}
\affiliation{%
    \institution{Beijing University of Posts and Telecommunications}
    \city{Beijing}
    \country{China}
}
\email{zhaokang@bupt.edu.cn}

\author{Jianwang Zhai}
\authornote{Corresponding author. This work is supported by the National Natural Science Foundation of China (No.~62404021), the Beijing Natural Science Foundation (No.~L2603021), and the Fundamental Research Funds for the BUPT (No.~2025AI4S20).}
\orcid{0000-0002-1581-3536}
\affiliation{%
    \institution{Beijing University of Posts and Telecommunications}
    \city{Beijing}
    \country{China}
}
\email{zhaijw@bupt.edu.cn}

\renewcommand{\shortauthors}{Jia et al.}

\begin{abstract}

Large language model (LLM) inference during autoregressive decoding is fundamentally memory-bandwidth-bound. 
Existing three-dimensional (3D) stacked processing-in-memory (PIM) architectures alleviate the off-chip bandwidth wall, yet remain constrained by several limitations: physical layer (PHY), controllers, and direct memory access (DMA) engines compete with arithmetic units for advanced-node base-die area; hybrid-bonding keep-out zones fragment the base-die floorplan and impede physical design; compute tiles built in a dynamic random access memory (DRAM) process are restricted to low logic density, capping achievable throughput; and nonlinear operations offloaded to the host reintroduce a communication bottleneck. 

We present Tri-Stack-PIM, a heterogeneous three-die architecture that resolves these tensions by dedicating a top DRAM die to storage, a middle buffer die on a mature process node to PHYs, DMA engines, and near-memory compute units, and a bottom base die on an advanced node exclusively to single-instruction-multiple-data (SIMD) multiply-accumulate (MAC) arrays. 
We further introduce a Unified Buffer-Aware Ready-based (UBAR) hardware out-of-order micro-operation scheduler that enforces cross-die dependency correctness at runtime while maximizing issue-slot utilization through a multi-term priority function, recovering up to 90\% of the cycles that an in-order policy would forfeit. 
A joint architecture-dataflow co-exploration identifies Pareto-optimal configurations that reduce the energy-delay product (EDP) by 33-66$\times$ over na\"ive parameter selection. 
Evaluated across multi-head, grouped-query, and multi-query attention (MHA, GQA, MQA) mechanisms with batch sizes 1-32, Tri-Stack-PIM achieves geometric-mean speedups of 4.95$\times$, 2.57$\times$, and 1.68$\times$ across all three attention mechanisms and batch sizes over a graphics processing unit (GPU) baseline, AttAcc, and H$^2$-LLM, respectively.

\end{abstract}



\maketitle

\section{Introduction}
\label{sec:introduction}

LLMs are now deployed across a wide range of applications, such as conversational assistants, code generation, document summarization, retrieval-augmented question answering, and agentic workflows. 
During autoregressive decoding, generating each token requires fetching the full set of model weights along with a key-value (KV) cache that grows with sequence length; as a result, model throughput is mainly bound by memory bandwidth.

3D-stacked PIM is one promising solution to this bottleneck, and two complementary hardware trends make it practical. 
First, near-memory and in-memory computing move bandwidth-sensitive computation adjacent to—or directly inside—the DRAM banks, eliminating off-chip traffic and exploiting the abundant internal bandwidth of the memory array~\cite{kim2023samsung,park2024attacc,heo2024neupims,li2025h2,kwon20221ynm,ke2021axdimm,gomezluna2022upmem}. 
Second, hybrid bonding (HB) provides dense, low-parasitic die-to-die interconnects that deliver far higher inter-die bandwidth and lower transfer energy than conventional microbump-based stacking, allowing logic and memory dies to be integrated at fine pitch~\cite{wuu20223d,lu2024hybrid,wang2023135,fujun2020stacked}. 
Together, these two advances enable practical 3D-stacked architectures for LLM decoding.

Prior LLM-oriented PIM systems have explored this direction along several paradigms. 
First, DRAM-process PIM~\cite{park2024attacc,kwon20221ynm} places compute logic directly inside the DRAM array. 
This maximizes proximity to the data, but is fundamentally constrained by the low logic density achievable in a DRAM process, which limits the arithmetic capability that can be integrated. 
Second, microbump-based 3D-stacked designs~\cite{heo2024neupims} typically retain operations other than general matrix multiplication (GEMM) on the host processor. 
This reintroduces host-device data movement, so performance is again bottlenecked by the bandwidth-limited host-device link that PIM is meant to eliminate. 
Third, HB has recently matured into a practical die-to-die integration technology~\cite{he2026tau,kagawa2016hybridbonding,beyne2024w2w}, enabling a more flexible partitioning. HB-based 3D-stacked systems~\cite{wuu20223d,li2025h2,li2026hardware} stack a dedicated compute die—fabricated in a high-performance logic process—directly beneath the DRAM die, and connect the two through dense HB interconnects that offer high inter-die bandwidth. 
Decoupling the compute die from the DRAM process removes the logic-density limitation of the first paradigm, while the HB link avoids the host-device bottleneck of the second. 
This flexibility, however, comes at a cost: the HB bonding pads impose keep-out zones on the base die, fragment the logic die into many small regions, increasing placement and routing complexity for compute logic~\cite{li2026hardware}.

\begin{table}[!t]
\centering
\caption{Comparison of existing designs and our Tri-Stack-PIM.
Columns indicate support for advanced-node compute logic
(Adv.-Node Compute), in-stack softmax execution (In-Stack Softmax),
compute/interface physical decoupling (C/I Decoupling), and
hardware-dynamic micro-operation scheduling (Dyn.\ Sched.).}
\small
\setlength{\tabcolsep}{4pt}
\begin{tabular}{lcccc}
\toprule
Work & Adv.-Node & In-Stack & C/I & Dyn. \\
     & Compute   & Softmax  & Decoupling & Sched. \\
\midrule
\multicolumn{5}{l}{\textit{DRAM-process PIM}} \\
TransPIM~\cite{zhou2022transpim}   & \xmark & \cmark & \xmark & \xmark \\
AttAcc~\cite{park2024attacc}       & \xmark & \cmark & \xmark & \xmark \\
NeuPIMs~\cite{heo2024neupims}      & \xmark & \xmark & \xmark & \xmark \\
IANUS~\cite{seo2024ianus}          & \xmark & \xmark & \xmark & \xmark \\
SpecPIM~\cite{li2024specpim}       & \xmark & \xmark & \xmark & \cmark \\
\midrule
\multicolumn{5}{l}{\textit{Logic-die / HB-based 3D-stacked}} \\
Duplex~\cite{yun2024duplex}        & \cmark & \cmark & \xmark & \cmark \\
H$^2$-LLM~\cite{li2025h2}          & \cmark & \xmark & \xmark & \xmark \\
\midrule
\textbf{Tri-Stack-PIM} & \textbf{\cmark} & \textbf{\cmark}
                       & \textbf{\cmark} & \textbf{\cmark} \\
\bottomrule
\end{tabular}
\label{table1}
\end{table}

As Table~\ref{table1} shows, no prior design closes all four gaps at once. DRAM-process PIM is capped by the low logic density of the DRAM process; host-side handling of nonlinear (non-GEMM) operations reintroduces the bandwidth-limited host-device bottleneck that PIM should remove; interface logic and control remain interleaved with arithmetic on the same compute-dedicated die; and none supports dynamic, cross-die scheduling at runtime. This last coupling is the most costly. When wide HB interfaces, host-facing PHYs, and control share one logic die with dense compute engines, they compete for the advanced-node silicon and routing meant for arithmetic. The cost is twofold: the consumed silicon, and the keep-out zones that HB bonding pads impose on the base die—fragmenting routable area and tightening place-and-route. This pressure only intensifies as interface width and operator complexity scale. Closing these gaps is not a matter of incremental refinement—a better neural processing unit (NPU) microarchitecture or a faster host link leaves the underlying coupling intact. What the industry needs is a fundamentally new, forward-looking 3D architecture.

We present \emph{Tri-Stack-PIM}, a heterogeneous three-die 3D-stacked PIM architecture that resolves the above bottlenecks by physically disaggregating storage, interfacing, and computation across three specialized dies. A top DRAM die provides high-capacity storage; a middle Buffer Die on a mature node handles interfacing, control, and nonlinear computation; and a bottom Base Die on an advanced node is reserved exclusively for dense matrix computation. The Buffer Die is the key enabler, and it pays off in two distinct ways. First, it restores a clean floorplan for the Base Die. To sustain full bandwidth, HB partitions the DRAM die into hundreds of independently-accessed banks; bonding their input/output (I/O) pads directly to the compute die below would scatter bond points across its entire area, shattering the floorplan into small, disjoint regions and sharply inflating placement-and-routing complexity for the compute logic~\cite{li2026hardware}. Inserted between the two, the Buffer Die aggregates the per-bank I/Os and re-exposes them through a regularized interface, leaving the Base Die a contiguous canvas for arithmetic—at negligible cost to delivered bandwidth. Second, it turns the freed mature-node area into an active execution layer. Aggregation consumes only a fraction of the Buffer Die—so what should fill the rest? We use it to pull every fixed-function block off the critical compute die: the host-facing PHYs, the HB controller, the DMA engines, the softmax and reduction engines (SREs) that execute nonlinear operators, and the channel-local dynamic scheduler. This is a deliberate technology-function match. High-speed I/O, control, and datapath glue scale poorly with process node, so hosting them on a mature die costs almost nothing in performance while reclaiming scarce advanced-node silicon for the one thing that truly rewards a leading-edge process: compute density. Better still, co-locating the nonlinear operators here keeps all non-GEMM work inside the stack, eliminating the host round-trips that throttle prior designs.

The main contributions of this work are:
\begin{itemize}[leftmargin=*]
\item We show that coupling interface, control, and compute on a single logic die throttles compute provisioning as HB width scales—both by consuming advanced-node silicon and by fragmenting the base die's floorplan—and propose a heterogeneous three-die partition that separates DRAM storage, mature-node interface/control, and advanced-node dense compute.
\item We design a UBAR-based scheduler on the Buffer Die that tracks dependency readiness, DRAM timing, and backend occupancy to issue micro-operations out of order, reclaiming cross-die idle cycles that static scheduling cannot.
\item We develop a coupled search over dataflow choices and hardware parameters that finds markedly better-balanced operating points than tuning either in isolation.
\item Across the evaluated workloads, Tri-Stack-PIM achieves geometric-mean speedups of $4.95\times$ over GPU-only, $2.57\times$ over AttAcc, and $1.68\times$ over H$^2$-LLM.
\end{itemize}

\section{Motivation}
\label{sec:motivation}

\begin{figure}[t]
  \centering
  \includegraphics[width=\linewidth]{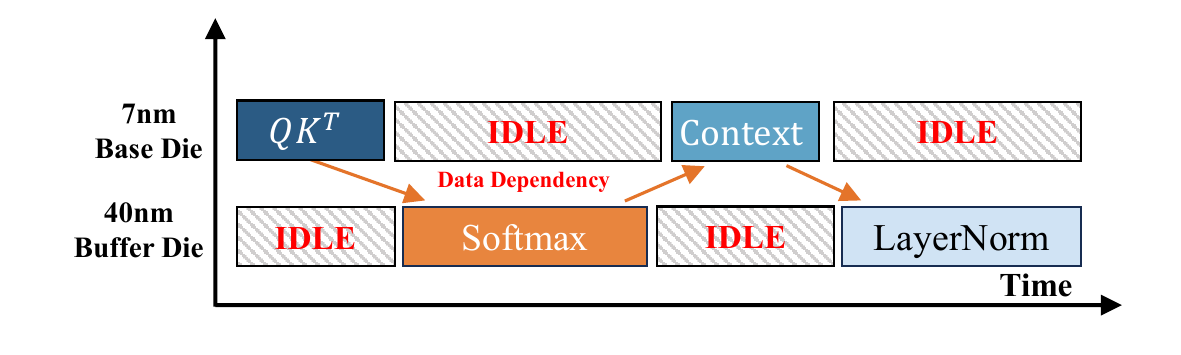}
  \caption{Per-channel execution trace of attention in Tri-Stack-PIM.}
  \label{fig:attention_execution_timeline}
\end{figure}

\begin{figure*}[t]
    \centering
    \includegraphics[width=\textwidth]{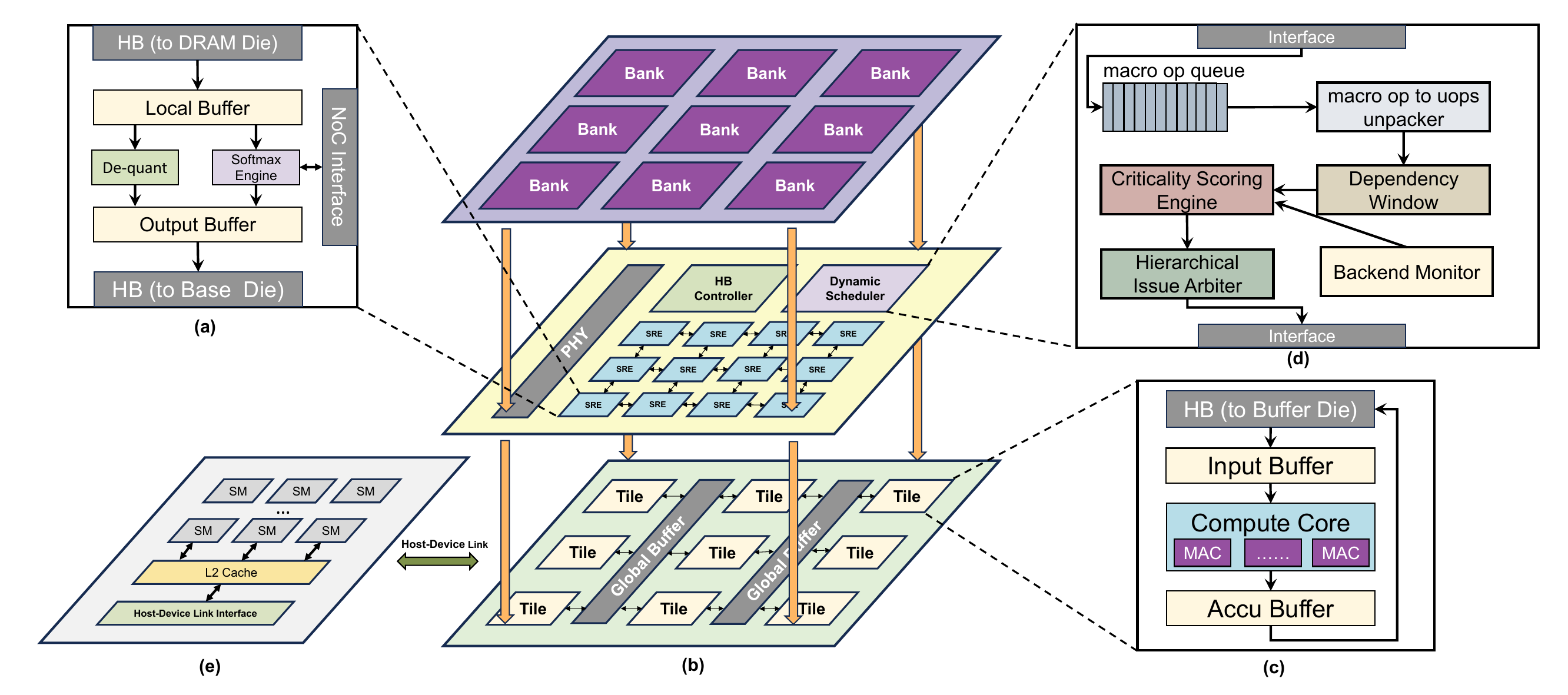}
    \caption{Tri-Stack-PIM overview. The top DRAM die stores weights and KV cache. The middle Buffer Die hosts host-facing PHY, HB controller, DMA, dynamic scheduler, and SREs. The bottom Base Die concentrates advanced-node silicon on compute tiles with local static random access memory (SRAM) buffers.}
    \label{overview}
\end{figure*}

As planar scaling yields diminishing returns~\cite{he2026tau}, 3D integration has become the primary architectural path for sustaining performance growth. Within this paradigm, HB constitutes the most aggressive realization: by transcending the pitch constraints of conventional microbumps~\cite{zhou2024research, lu2024hybrid}, HB achieves sub-micron interconnect density that unlocks vertical memory bandwidth at a granularity fundamentally unattainable through prior 3D stacking approaches—transforming the economics of memory-bound computation.

However, state-of-the-art hybrid-bonded 3D PIM architectures remain subject to several practical constraints. First, because each DRAM bank bonds vertically to the base die through a dedicated HB interface, the proliferation of banks induces a commensurate proliferation of keep-out zones on the logic tier, progressively fragmenting the floorplan and penalizing placement-and-routing closure. Second, the co-integration of host-facing high-speed I/O, and control infrastructure consumes non-trivial silicon area on the base die, eroding the budget available for compute logic.

From an algorithmic standpoint, the arithmetic demand imposed on the logic die during decoding is intensifying. The widespread adoption of GQA/MQA~\cite{ainslie2023gqa, shazeer2019fast} means that a single KV-cache fetch is mathematically broadcast across $G$ query heads (e.g., $G{=}8$ in LLaMA-2-70B), while batched inference further amplifies effective arithmetic intensity in proportion to $G \times B$, where $B$ denotes batch size. Exploiting this reuse, however, demands a dedicated on-die buffer to stage KV heads across the broadcast and batching dimensions. Consequently, surrendering an excessive fraction of the already-constrained logic-die silicon to non-compute structures directly strangles the compute-engine budget and forfeits the very arithmetic density that modern attention mechanisms could otherwise sustain.

Consequently, existing HB-enabled PIM systems encounter a fundamental tension: the logic die is squeezed simultaneously from both the physical and algorithmic directions—per-bank keep-out zones and host-facing I/O infrastructure fragment and consume its already finite silicon, while GQA/MQA broadcast reuse ($G \times B$ arithmetic intensity) demands precisely the dense compute fabric that such overheads displace.

These limitations motivate our heterogeneous three-die stacking strategy. The maturation of multi-layer hybrid bonding and fine-pitch through-silicon-via (TSV) interconnects in industrial packaging~\cite{li2026hardware,he2026tau,wang2023135,wang2025_3duam} now renders such multi-tier 3D integration manufacturable. We leverage these advances to interpose a mature-node buffer die between the compute and DRAM tiers. The benefits are twofold. First, redistributing all DRAM-facing hybrid-bond I/Os onto the buffer tier—rather than the logic die—eliminates the per-bank keep-out zones that would otherwise fragment the compute floorplan and inflate placement-and-routing complexity. Second, what should occupy the residual buffer-die area? We make the decision to co-locate the high-speed I/O PHYs, the nonlinear-acceleration unit, and controllers on this die. The rationale is economic: analog/mixed-signal circuits gain little from dimensional scaling, and advanced FinFET nodes actively worsen noise, variability, and voltage-headroom margins for PHY designs~\cite{razavi2002cmos}. A mature planar node (e.g., 40/45\,nm) thus provides a superior substrate—lower cost, simpler process, well-characterized models—while liberated advanced-node silicon is reclaimed exclusively for computation-heavy engines.

We therefore propose Tri-Stack-PIM (as shown in \Cref{fig:attention_execution_timeline}), a heterogeneous three-die stacked architecture: the top DRAM dies furnish storage capacity; the sandwiched mature-node buffer die hosts high-speed I/O PHYs and memory-proximal low-arithmetic operators; and the bottom advanced-node base die dedicates its silicon exclusively to dense arithmetic.


However, this architecture introduces a critical scheduling challenge. Because the Base Die embodies the costliest advanced-node silicon in the stack, its utilization is the first-order determinant of attention throughput. Yet strict data dependencies between Base-Die GEMMs and Buffer-Die Softmax inherently expose the pipeline to Base-Die stalls absent effective scheduling (\Cref{fig:attention_execution_timeline}).

A static software schedule is the first step: a compiler constructs a macro-operation directed acyclic graph (DAG), determines operator placement, and overlaps independent operators across dies~\cite{dennis1974preliminary, fisher1983very, nikhil1990executing}. 
This macro-instruction granularity marks the exact boundary of compile-time schedulability. Yet each macro-instruction decodes into numerous micro-operations whose readiness depends on runtime state-DRAM timing, die-to-die link congestion, heterogeneous-unit availability-none statically predictable. A static schedule is therefore brittle, and streaming compiler-generated µop sequences over the host link would trigger a control-traffic explosion. Inspired by the CDC 6600 scoreboard~\cite{thornton1964parallel}, we adopt a hardware-dynamic out-of-order µop scheduler that resolves these hazards at runtime.

Moreover, µops contend simultaneously for heterogeneous resources; mere hazard detection for correctness alone is insufficient without scheduling quality. The scheduler must perform prioritized multi-resource arbitration that maximizes utilization. We therefore introduce UBAR, a micro-scheduler that couples out-of-order issue with quality-driven multi-resource arbitration to dynamically orchestrate µop execution.

In summary, we propose Tri-Stack-PIM—a heterogeneous three-die architecture interposing a mature-node buffer die between DRAM and an advanced-node compute die to consolidate I/O and mixed-signal functions off-chip while reclaiming compute silicon exclusively for dense arithmetic—and UBAR, a hardware-dynamic out-of-order micro-scheduler coupling hazard resolution with quality-driven multi-resource arbitration to orchestrate µop execution where static scheduling cannot.

\section{Tri-Stack-PIM Architecture}
\label{sec:methodology}

\begin{figure*}[t]
  \centering
  \includegraphics[width=\linewidth]{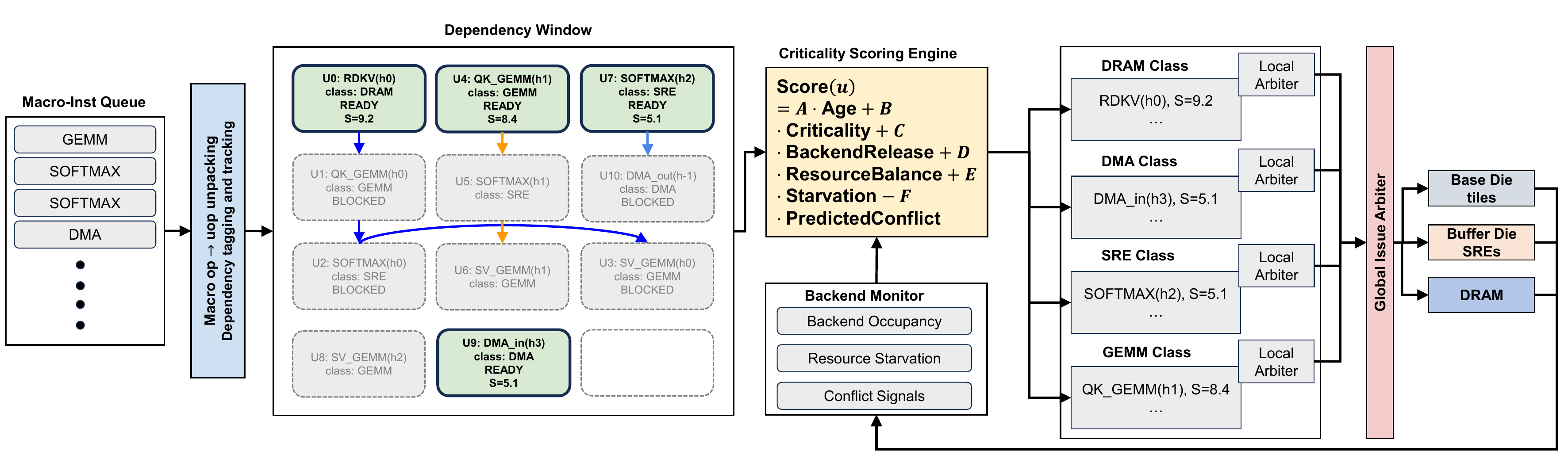}
  \caption{Dynamic scheduling pipeline on the Buffer Die: software emits macro-instructions that are unpacked into dependency-tracked micro-operation windows; a two-stage arbiter selects ready micro-operations using the UBAR priority score computed from runtime counters.}
  \label{fig:fig7}
\end{figure*}

\subsection{Design Principle}
Tri-Stack-PIM is a command-driven dataflow near-memory platform.
It contains no program counter, no instruction cache, and no branch predictor; execution advances by dependency readiness within an offline-compiled DAG. 
This maximizes arithmetic silicon on the base die. 
Three principles underlie the design. (1)~\textit{Data Proximity.} Rather than incrementally optimizing GPU/NPU architecture, Tri-Stack-PIM attacks the memory wall at its root: each DRAM bank connects via hybrid bonding directly to a dedicated compute tile, delivering ultra-high bandwidth at ultra-low latency. (2)~\textit{Separation of Compute and Control.}
In 2 die stacked designs, host-facing I/O, controllers, and scheduling logic compete for area and routing resources with the compute fabric.
Tri-Stack-PIM resolves this by concentrating all non-compute functions onto the buffer die (\Cref{overview}), leaving the base die's entire area budget for arithmetic. This yields two advantages. First, the buffer die is fabricated on a mature node well suited to mixed-signal PHYs and control datapaths, while the base die targets an advanced node that maximizes MAC density per unit area. Second, it eliminates routing congestion on base die. (3)~\textit{Hybrid Static/Dynamic Scheduling.} Programming this near-memory platform presents a fundamental tension:
a multiple-instruction multiple-data (MIMD) model demands per-tile firmware of prohibitive complexity;
a single-instruction multiple-thread (SIMT) model requires a full compiler backend mapping attention kernels onto lockstep lanes.
Tri-Stack-PIM sidesteps both extremes with a clear division of scheduling responsibility.
The host compiler owns macro-level decisions---emitting a statically scheduled DAG of macro-instructions that encodes operator type, operand placement, and inter-operator dependencies---while UBAR resolves all µop-level ordering at runtime, exploiting dynamic state invisible to any compile-time analysis.

Figure~\ref{overview} illustrates the overall architecture.
The top DRAM die provides storage; the bottom base die contains one compute tile per bank—each equipped with SIMD lanes—mapped one-to-one to the DRAM banks above via hybrid bonds. The middle buffer die integrates the host-facing PHY, UBAR scheduler, DMA engine, and SRE units for near-memory softmax and reduction.

\subsection{Execution Model}

\subsubsection{Hierarchical Instruction Set Architecture.}

The host compiler emits macro-instructions, each encoding a coarse-grained
operation together with its data-dependency constraints. Four macro types
suffice to express the complete attention dataflow:
(1)~\texttt{LAUNCH\_GEMM}---a tiled matrix--vector multiply distributed across
DRAM banks and base-die compute tiles, carrying a \texttt{repeat\_count} field
that encodes the GQA query-head fan-out over the fetched KV tile;
(2)~\texttt{SRE\_SOFTMAX}---softmax accumulation and partial-sum reduction
executed in the buffer-die SRE;
(3)~\texttt{DMA\_H2D}---host-to-device tile transfer;
(4)~\texttt{MODE\_SWITCH}---a global barrier that serializes execution phases.
Every macro carries a compiler-assigned tag together with a predecessor list
(\texttt{pre\_tag}), collectively forming a static inter-macro dependency DAG;
a successor may issue only after all of its predecessors have retired.
Because macros belonging to independent GQA groups carry no cross-group
dependencies, they may be dispatched concurrently across channels.

Upon arrival at the per-channel controller, each macro-instruction is decoded
into micro-operations (\textmu{}ops) and posted to a hardware scoreboard. The
controller maintains no program counter: issue eligibility is purely
predicate-driven---a \textmu{}op becomes ready when its outstanding dependency
count reaches zero and the requisite execution resources are available.

\subsubsection{Attention Dataflow.}

A single attention-head computation traverses four stages.
\textcircled{1}~The controller issues \texttt{ACTAB}/\texttt{RDAB}
commands (All-Bank Activate and Read-All-Banks) to the DRAM die; each
column burst delivers a KV-cache tile through the hybrid-bond interface,
throttled by DRAM timing ($t_\mathrm{RCD}$, $t_\mathrm{CCD,AB}$,
$t_\mathrm{RP}$, refresh).
\textcircled{2}~The tile lands in a base-die compute tile's local SRAM
and is consumed by an SIMD MAC array; compute latency per burst is
$\lceil (W_\mathrm{HB}/b_\mathrm{quant} \times r) / N_\mathrm{lanes}
\rceil$ cycles, where $W_\mathrm{HB}$ is the HB data width,
$b_\mathrm{quant}$ the quantization width, $r$ the GQA repeat count,
and $N_\mathrm{lanes}$ the SIMD lanes.
\textcircled{3}~Partial query-key (QK) dot-product results are forwarded upward
through the HB interface to per-bank SREs on the buffer die; when the
head's hidden dimension is sharded across multiple banks, each SRE
relays its partial scores over an on-die network-on-chip (NoC) to a geometrically elected
master SRE for cross-bank reduction prior to softmax.
\textcircled{4}~The master SRE executes softmax in a single streaming
pass---doing maximum, exponential rescaling, and denominator
accumulation. The normalized attention weights are then forwarded back
through the HB interface to the base-die compute tiles, which execute a
second \texttt{LAUNCH\_GEMM} phase---fetching V-cache tiles from DRAM
and accumulating the weighted value sum---to produce the final per-head
context vector.

\subsubsection{Base-Die Compute Tile}

Each compute tile is purely data-triggered (\Cref{overview}c). It
comprises three structures: (1)~a local SRAM buffer absorbing incoming
bursts, with backpressure to the scheduler when full; (2)~an
$N_\mathrm{lanes}$-wide fully pipelined SIMD MAC array processing one
element per lane per cycle, serving both the QK and score-value (SV)
\texttt{LAUNCH\_GEMM} phases sequentially; and (3)~a ready-tick register that serializes successive bursts to the
same tile, enforcing the constraint that the MAC array cannot accept a
new chunk until the prior one retires.

\subsubsection{Buffer-Die SRE.}

As shown in \Cref{overview}(a), the SRE resides on the buffer die, where the mature node provides ample area at low cost and eliminates the $>$1\,\textmu s host round-trip that would otherwise be incurred by offloading softmax to the host. One SRE is instantiated per bank, with all SREs within a channel interconnected by a 2D-mesh NoC. Each SRE comprises a local SRAM buffer for staging partial-result tiles arriving from the base die, a reduction unit for cross-bank accumulation, and a softmax unit that performs running-max tracking, exponential rescaling, and denominator accumulation in a single streaming pass. Execution proceeds in four phases: \textcircled{1}~base-die tiles traverse the hybrid-bonding interface into each co-located SRE buffer; \textcircled{2}~when cross-bank reduction is required, a master SRE elected at the geometric centroid gathers tiles across the mesh (per-hop latency: 2 cycles; total transfer cost $\lceil\mathit{tile\_bytes}/\mathit{BW}_\mathrm{NoC}\rceil + h \cdot t_\mathrm{hop}$); \textcircled{3}~the master accumulates the gathered partials into a single score vector; \textcircled{4}~softmax normalizes the result and writes the attention weights back for the subsequent SV-GEMM.

\subsection{UBAR Dynamic Scheduler.}
A single macro-instruction decomposes into tens to hundreds of µops whose completion latencies depend on runtime information that are fundamentally unpredictable at compile time. A naïve in-order policy enforces strict head-of-line ordering: whenever the oldest ready µop is denied issue due to a timing or resource constraint, all younger µops stall irrespective of their individual eligibility, squandering a substantial fraction of available issue slots under high concurrency. To overcome this limitation, we adopt out-of-order µop scheduling. To further improve issue quality, UBAR incorporates a priority-scoring mechanism.

Tri-Stack-PIM employs a two-level abstraction (\Cref{fig:fig7}). The host emits coarse-grained macro-instructions with compile-time dependency edges (\texttt{tag}/\texttt{pre\_tag}); a per-channel hardware unpacker expands each macro into fine-grained µops posted on a scoreboard with predecessor counts. A µop becomes ready when its dependency count reaches zero, entering the scheduling window visible to the arbiter.

The UBAR scheduler assigns each ready µop $u$ a composite priority score:
\begin{equation}
  S(u) = \alpha\,T_\text{age} + \beta\,T_\text{crit} + \gamma\,T_\text{back} +
  \delta\,T_\text{bal} + \epsilon\,T_\text{starv} - \zeta\,T_\text{conf},
  \label{eq:ubar}
\end{equation}
where $T_\text{age}$ increases linearly with residence time and saturates at a configurable horizon, thereby favoring older µops without the unbounded growth that would reduce the scheduler to first-in-first-out (FIFO) order.
$T_\text{crit}$ encodes a static critical-path class assigned at µop creation time (Barrier\,$>$\,Softmax\,$>$\,GEMM\,$>$\,DRAM\,$>$\,DMA).
$T_\text{back}$ monitors the fill ratio of the base-die GEMM input FIFO and activates only when occupancy drops below a critical threshold, elevating the priority of DRAM-read µops capable of replenishing the data-starved compute pipeline---this constitutes the primary mechanism for breaking head-of-line blocking, and accordingly carries the largest default weight ($\gamma = 5\times$ the baseline).
$T_\text{bal}$ tracks per-class cumulative issue counts and penalizes over-represented classes, preserving issue diversity across heterogeneous resource types.
$T_\text{starv}$ provides an independent anti-starvation guarantee: it remains identically zero until a patience threshold is exceeded, then ramps to full scale at a saturation bound, ensuring bounded worst-case latency through a mechanism distinct from the smooth aging of $T_\text{age}$.
$T_\text{conf}$ is the sole subtractive term; it queries the DRAM timing model's per-bank open-row state and applies a full-scale penalty to row-buffer-missing DRAM-access µops, steering issuance away from predicted bank conflicts.

As shown in \Cref{fig:fig7}, each channel issues µops through a two-stage arbiter designed to split the critical selection path for timing closure. In the first stage, ready µops are grouped into four priority classes and the highest-scoring entry in each class is elected as class champion; in the second stage, a global argmax selects the overall winner from among the four champions, falling through to the next-ranked champion on a resource-check failure rather than stalling. This decomposition bounds the selection fan-in to four regardless of scoreboard depth and closes the arbiter within a single buffer-die cycle.

\subsection{Joint Architecture–Dataflow Co-Exploration (DSE)}
\label{dse}
The design parameters are tightly coupled: widening the hybrid-bonding I/O increases the per-read data volume yet yields no speedup once the SIMD array ceases to be memory-bound; conversely, widening the SIMD array elevates peak throughput only if the dataflow and workload characteristics keep enough tiles in flight to saturate it---that is, only if sufficient parallelism is exposed. Because no parameter can be tuned in isolation, a joint co-exploration of architecture and dataflow is necessary.

Table~\ref{tab:dse_space} enumerates seven parameters spanning four domains. Notably, the dataflow choice denotes the head-to-bank mapping, which determines how attention heads are distributed across DRAM banks and channels, thereby governing both inter-bank parallelism and row-buffer reuse. The arithmetic intensity of the attention kernel shifts with the group ratio $G = N_\text{heads}/N_\text{KV}$: MHA ($G{=}1$) is predominantly memory-bound, MQA ($G{=}16$) is predominantly compute-bound, and GQA ($G{=}4$) stresses both axes simultaneously. We therefore explore three representative models spanning these regimes---OPT-6.7B (MHA), PaLM-8B (MQA), and LLaMA3-8B (GQA). We employ multi-objective Bayesian optimization via Ax/BoTorch~\cite{balandat2020botorch, olson2025ax}, jointly minimizing latency, die cost, and power. Hardware block sizes and power modeling are detailed in \Cref{experiment_setup}. The resulting Pareto fronts are analyzed in \S\ref{sec:evaluation}.

\begin{table}[t]
\centering
\caption{Joint design space. Each parameter interacts with at least one parameter in a different domain.}
\label{tab:dse_space}
\resizebox{\columnwidth}{!}{%
\begin{tabular}{l l l}
\toprule
\textbf{Domain} & \textbf{parameters} & \textbf{Choices} \\
\midrule
Dataflow      & KV bank mapping        & Round-robin, sequence-shard $\times\{1,2,4,8\}$ \\
\midrule
Base Die      & SIMD lanes ($N_\text{lanes}$)   & \{32, 64, 128, 256, 512\} \\
              & Clock frequency                  & 0.5–1.5\,GHz (11 steps) \\
              & Input buffer capacity            & 32\,KB–8\,MB (7 steps, log-spaced) \\
\midrule
Buffer Die    & Clock frequency                  & 0.1–1.0\,GHz (10 steps) \\
              & SRE buffer capacity              & 32\,KB–8\,MB (7 steps, log-spaced) \\
\midrule
Interconnect  & HB data width ($W_\text{HB}$)   & \{128, 256, 512, 1024, 2048, 4096\}\,bits \\
\bottomrule
\end{tabular}%
}
\end{table}

\section{Evaluation}
\label{sec:evaluation}

\begin{figure*}[t]
\centering
\includegraphics[width=\linewidth]{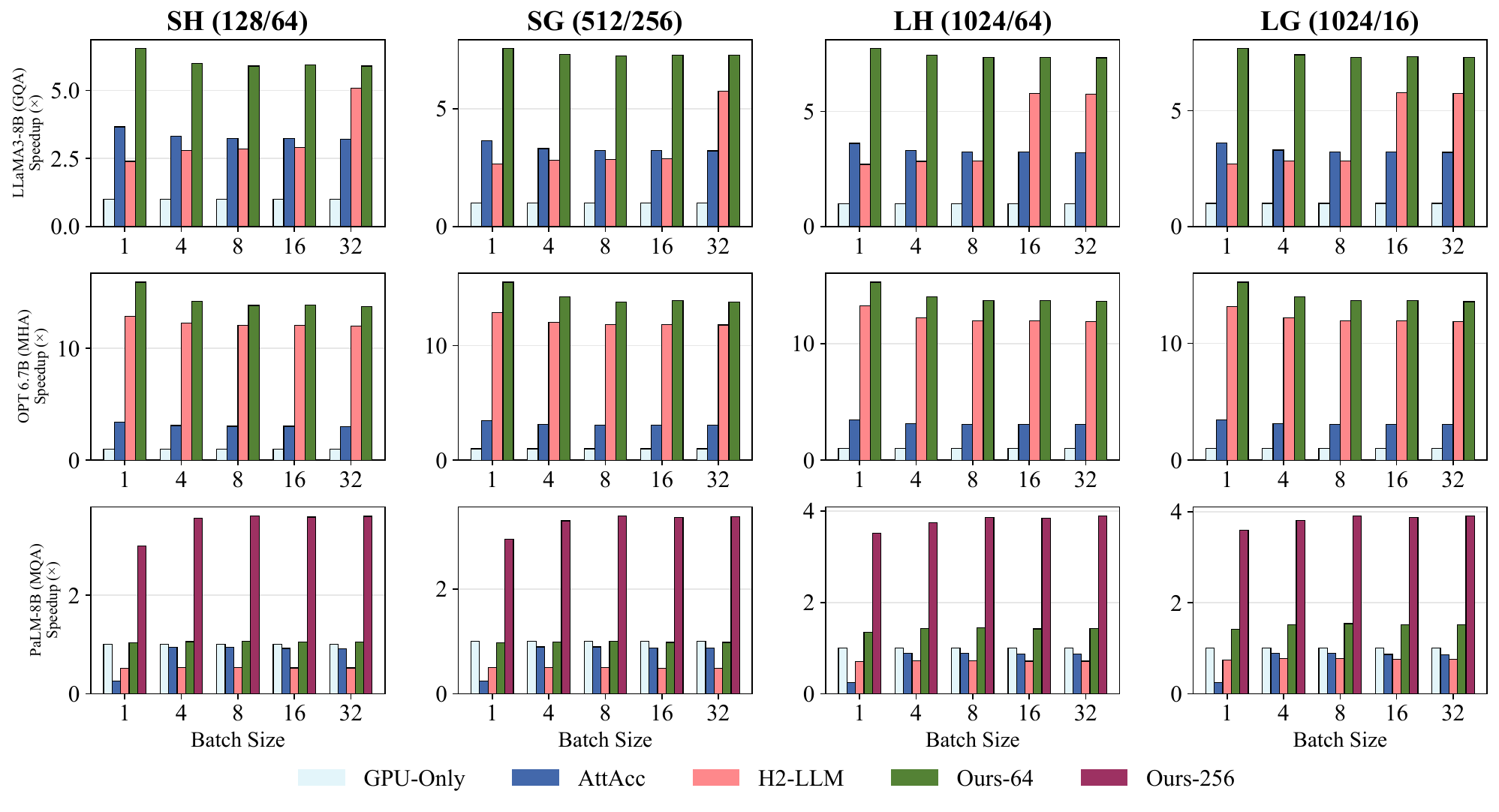}
\caption{Per-token attention speedup across all evaluated workloads. Tri-Stack-PIM consistently outperforms the GPU-only, AttAcc, and H$^2$-LLM baselines.}
\label{fig:fig8}
\end{figure*}

\subsection{Experimental Setup}
\label{experiment_setup}
We evaluate three models that span the attention-mechanism spectrum---OPT-6.7B (MHA, 32/32 heads), LLaMA3-8B (GQA, 32/8 heads), and PaLM-8B (MQA, 16/1 heads)~\cite{zhang2022opt,grattafiori2024llama,chowdhery2023palm}---across batch sizes $\{1,4,8,16,32\}$ and (prompt, generation)-length pairs $(128,64)$, $(512,256)$, $(1024,64)$, and $(1024,16)$. The Base-Die compute tile is synthesized with the ASAP7 process design kit (PDK)~\cite{clark2016asap7} (0.5--1.5\,GHz, 16--256 MACs/tile; 73.1--4614.5\,$\mu$m$^2$/block, 0.15--41.2\,fJ/MAC). The Buffer-Die SRE is synthesized with NanGate 45\,nm PDK~\cite{nangate45} (0.1--1.0\,GHz; 0.47\,mm$^2$, 1.1--9.5\,pJ/softmax op). On-chip SRAMs are modeled with FinCACTI~\cite{shafaei2014fincacti}: a 64\,KB buffer occupies 0.116\,mm$^2$ at 5.27\,pJ/access (7\,nm) versus 0.305\,mm$^2$ at 10.55\,pJ/access (45\,nm). Die-to-die link energy (0.88\,pJ/bit) is derived from the HB model of H$^2$-LLM~\cite{li2025h2}; Our Cycle-accurate simulation extends Ramulator~2~\cite{luo2023ramulator}.

We compare against three baselines. GPU-Only models a representative edge GPU backed by 4-channel LPDDR5X with 30.7\,GB/s effective bandwidth and 1.0\,TFLOPS effective FP16 compute, evaluated via the analytical model open-sourced by AttAcc~\cite{park2024attacc}. AttAcc~\cite{park2024attacc} is simulated cycle-accurately on the same LPDDR5X assumption using its bank-group-level PIM configuration (128\,GB/s internal bandwidth, 0.128\,TFLOPS). H$^2$-LLM~\cite{li2025h2} is evaluated on our cycle-accurate simulator under its published parameters mapped to the same LPDDR organization for a controlled comparison. All experiments are executed on an Intel Xeon Platinum 8375C server at 2.90\,GHz.

\subsection{Performance Analysis}
Since our work targets the attention mechanism rather than host-device orchestration, we report \emph{per-token average attention latency}---aggregated over all 32 decoder layers---as the metric. The base die operates at 1.3\,GHz and the buffer die at 1.0\,GHz uniformly across GPU-only, AttAcc, H$^2$-LLM, and Tri-Stack-PIM; unless otherwise specified, the default SIMD width is 64 lanes for Tri-Stack-PIM and 32 lanes for H$^2$-LLM. Figure~\ref{fig:fig8} presents the resulting attention latency across all workload configurations. Tri-Stack-PIM achieves geometric-mean speedups of $4.95\times$ over GPU-only, $2.57\times$ over AttAcc, and $1.68\times$ over H$^2$-LLM.

Across all configurations, Tri-Stack-PIM achieves geometric-mean speedups of $14.15\times$/$7.03\times$ over GPU-only, $4.50\times$/$2.12\times$ over AttAcc, and $1.16\times$/$2.05\times$ over H$^2$-LLM on the MHA (OPT-6.7B) and GQA (LLaMA3-8B) workloads, respectively. The MHA gains are expectedly larger: at $G{=}1$ the workload is strictly memory-bound, leaving the GPU baseline bandwidth-starved and maximizing the benefit of near-memory compute. A similar disparity appears against AttAcc, whose absence of base-die SRAM reuse incurs redundant DRAM accesses that penalize the memory-intensive MHA most acutely.

On the MQA workload (PaLM-8B), Tri-Stack-PIM (SIMD\,=\,256) achieves geometric-mean speedups of $3.58\times$ over GPU-only, $5.20\times$ over AttAcc, and $5.82\times$ over H$^2$-LLM. Three observations emerge. First, AttAcc underperforms the GPU in every configuration: serving 16 query heads through bank-group-level PIM without a reuse buffer incurs four redundant KV scans---one per bank group---whereas the GPU amortizes a single KV fetch via L2 reuse. Integrating on-die SRAM is precluded by the DRAM process, which lacks the metal stack required for dense 6T cells. Second, H$^2$-LLM also trails GPU-only: its host-offloaded softmax introduces a per-head synchronization barrier bounded by host-device bandwidth, and, absent cross-tier functional disaggregation, the base-die area constrains the SIMD datapath to 32 lanes---leaving the system compute-starved relative to available DRAM throughput. Third, Tri-Stack-PIM at 64 SIMD lanes falls below AttAcc because the MQA case is compute-bound; however, the base-die area budget permits scaling to 256 lanes, at which point Tri-Stack-PIM surpasses all baselines decisively.

\subsection{UBAR Dynamic Scheduler Effectiveness}

We isolate UBAR's contribution along two orthogonal axes: (a) the latency reduction from out-of-order issue alone---reordering \textmu ops beyond strict in-order---and (b) the incremental gain of UBAR's multi-term scoring function over a na\"ive age-priority out-of-order (OoO) baseline.

\subsubsection{Recoverable Scheduling Slack under In-Order Issue.}

\begin{figure}[t]
\centering
\includegraphics[width=\linewidth]{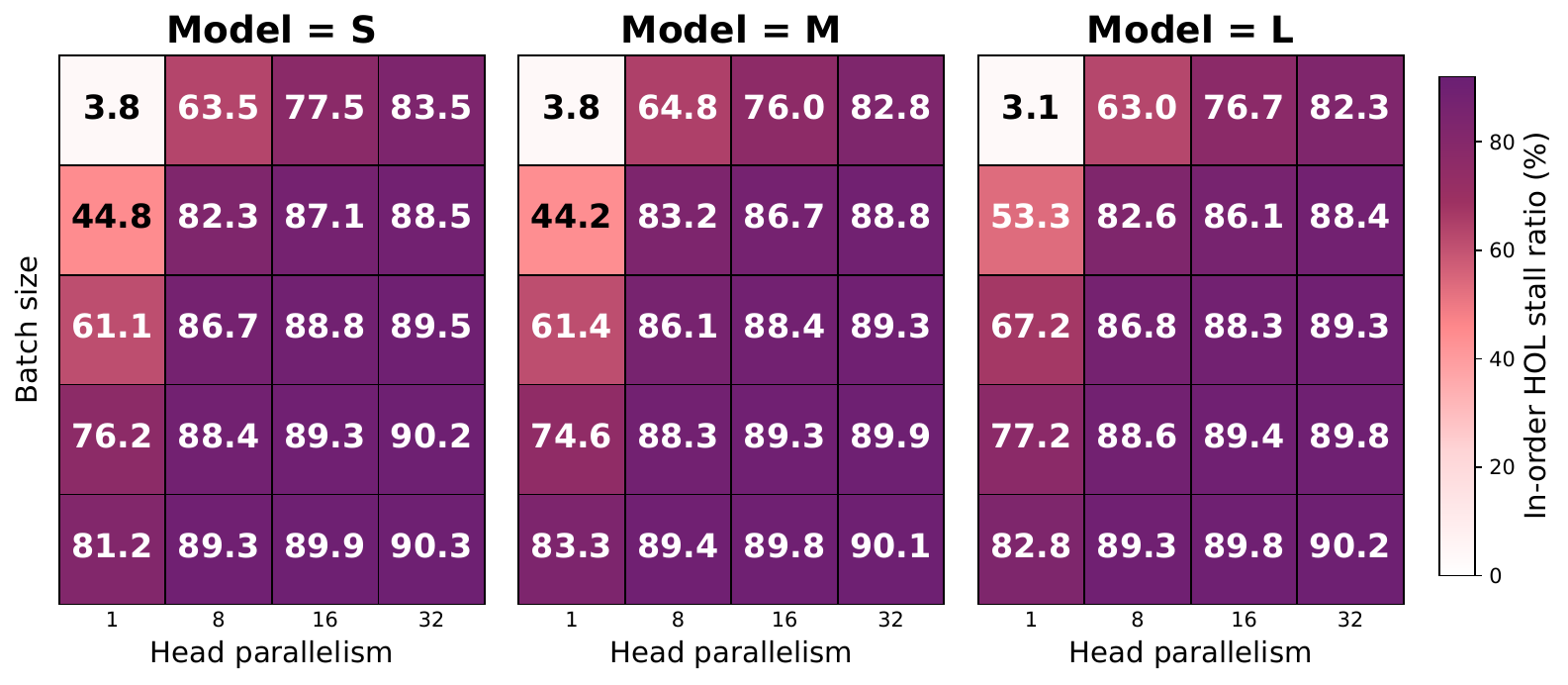}
\caption{Fraction of cycles where in-order issue would stall
on a blocked head uop despite a ready non-head uop being
available. This recoverable slack is what the UBAR dynamic
scheduler exploits.}
\label{fig:hol_heatmap}
\end{figure}

Figure~\ref{fig:hol_heatmap} quantifies the fraction of scoreboard-active cycles during which the In-Order stalls on a blocked head-of-line \textmu op while at least one ready \textmu op exists elsewhere.

Results find that head parallelism and batch depth independently create scheduling opportunity, and both saturate rapidly.
With $H{=}1$, $B{=}1$, the scheduler degenerates to a serial chain: only 3.5\% of active cycles exhibit recoverable slack.
Raising either axis alone suffices to unlock substantial opportunity: $H{=}1,\,B{=}8$ yields 63\% slack through cross-request interleaving; $B{=}1,\,H{=}8$ yields 64\% through intra-request head diversity.
Once both axes exceed a modest threshold ($B\!\ge\!8$, $H\!\ge\!8$), the head-of-line stall ratio saturates at 86--90\% regardless of model size, indicating that the in-order controller wastes nearly nine of every ten schedulable cycles on a conflict a ready-aware policy could bypass.
This near-universal saturation---insensitive to model size once concurrency is sufficient---establishes OoO issue not as a corner-case optimization but as an architectural necessity for any PIM controller serving multi-head batched inference.

\subsubsection{Contribution of UBAR's Multi-Term Scoring over OoO Alone.}

\begin{figure}[t]
\centering
\includegraphics[width=\linewidth]{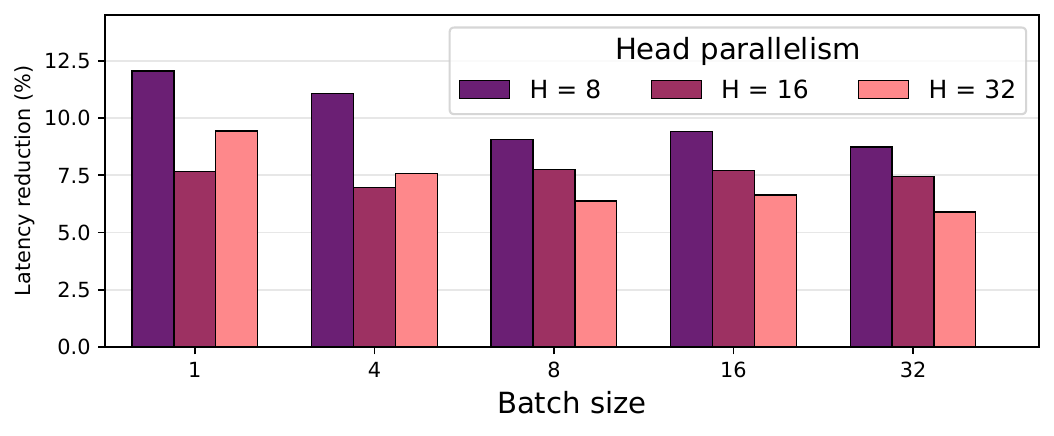}
\caption{Latency reduction of UBAR over age-first OoO,
grouped by batch size and head parallelism (averaged across model sizes
S/M/L). The dashed line marks the geometric mean (8.3\%).
Higher bank-conflict pressure (low $H$) amplifies the scoring advantage.}
\label{fig:fig_scoring_gain}
\end{figure}

To isolate the contribution of quality-driven arbitration beyond mere out-of-order issue, we compare UBAR against an age-first OoO baseline that zeroes all scoring weights except timestamp—effectively an oldest-ready-first policy. Both schedulers share identical hardware parameters (SIMD\,=\,32, HB\,=\,2048\,b, base die 500\,MHz, buffer die 1.0\,GHz, 1\,MB+4\,MB buffers) on a single-channel stress mapping with 8\,ch\,$\times$\,16\,banks. We sweep Model\,$\in$\,\{S(2k), M(4k), L(8k)\}, Batch\,$\in$\,\{1,4,8,16,32\}, and Heads\,$\in$\,\{8,16,32\} in MHA mode.

Figure~\ref{fig:fig_scoring_gain} reports UBAR's latency reduction over age-first across 32 configurations. Three observations emerge. (1)~\textit{Scoring outweighs reordering.} The geometric-mean scoring gain is 9.4\%, exceeding the 7.2\% from OoO reordering alone; na\"ive oldest-first captures less than half the total 17.2\% UBAR advantage. (2)~\textit{Gain peaks under low head parallelism.} At $H$\,=\,8 the mean scoring gain reaches 11.7\%, versus 8.2\% at $H$\,=\,32. (3)~\textit{Age-first can degrade performance.} At (S, $B$\,=\,1, $H$\,=\,8), age-first is 2.1\% slower than in-order first-come-first-served (FCFS).

\subsection{Design-Space Exploration Analysis}

We find that the Pareto-optimal sweet spots reduce attention latency by $28.5{\times}$, $17.1{\times}$, and $39.8{\times}$ for GQA, MHA, and MQA over the worst-case configuration, while compressing area by $130$--$95{\times}$ and power by $1.5$--$7.6{\times}$. The resulting $33$--$66{\times}$ EDP improvement confirms that the DSE simultaneously optimizes for speed and energy efficiency.

\textbf{Takeaway 1: Arithmetic intensity dictates the effective HB bandwidth requirement.}
Widening the hybrid bond from 128\,b to 4096\,b yields $12{\times}$ and $4.8{\times}$ median latency reductions for MHA and GQA, but only $2{\times}$ for MQA. MHA and GQA are memory-bound, so wider HB translates directly into proportional throughput gains. MQA, however, is compute-bound rendering additional bandwidth ineffective against this sequential bottleneck.
\textbf{Takeaway 2: Dataflow strategy must be co-designed with the attention pattern.}
At fixed HB width, mapping alone induces $2.3{\times}$ (MHA) to $9.6{\times}$ (MQA) latency variation. GQA favors contiguous hidden-dimension sharding to co-locate shared KV for concurrent multi-head access; MHA benefits from round-robin sequence sharding that distributes independent KV pairs across banks for maximum read parallelism; MQA prefers snake hidden-dimension sharding to spread computation across processing elements (PEs), exploiting SIMD parallelism on its single shared KV head. \textbf{Takeaway 3: SIMD lanes only pays off under compute-bound regimes.} For MQA, doubling SIMD lanes from 32 to 256 yields a near-linear $4.5{\times}$ speedup, as the single serialized KV head leaves ample arithmetic work per byte fetched. For MHA and GQA, wider SIMD provides negligible benefit ($<1.1{\times}$) because the bottleneck is DRAM bandwidth---widening the HB bus, not the datapath, is the correct lever.

\subsection{Overhead Analysis}
\label{sec:cost}

\begin{figure}[t]
\centering
\includegraphics[width=\linewidth]{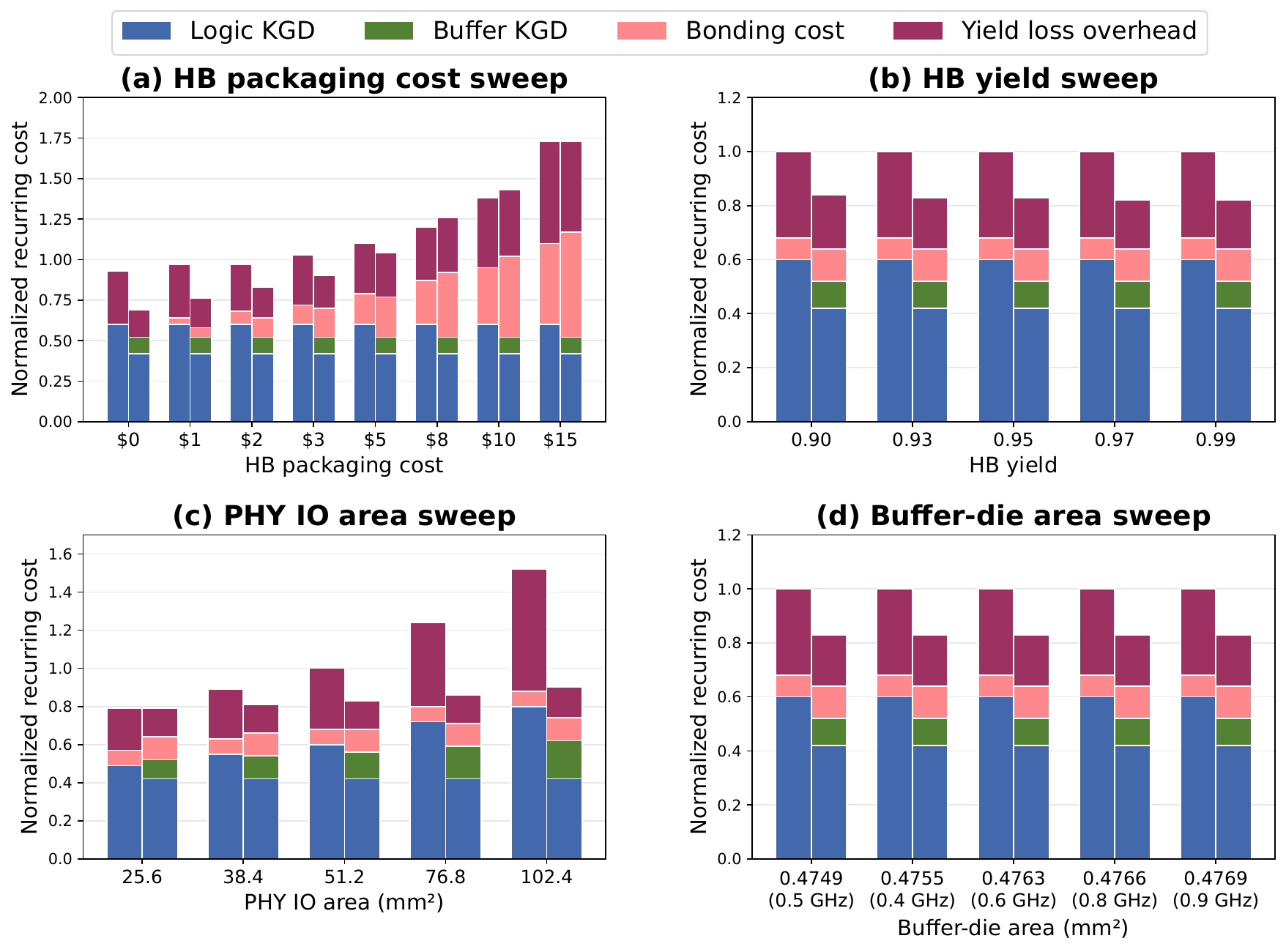}
\caption{Yield-aware recurring-cost breakdown of Tri-Stack-PIM versus a
two-die baseline, showing the tradeoff between reduced advanced-node logic
cost and the added Buffer-Die and bonding overhead.}
\label{fig:fig9}
\end{figure}

\noindent\textbf{Cost Analysis.}
We adopt the cost model of AccelStack~\cite{bai2025accelstack}.
Since the DRAM die is identical across both designs, it cancels in the
comparison; we report only recurring manufacturing cost, excluding
non-recurring engineering (NRE) expenditures.  The per-unit cost thus
reduces to
\begin{equation}
  C_{\mathrm{RE}}
    = \frac{C_{\mathrm{logic}}}{Y_{\mathrm{logic}}}
    + \frac{C_{\mathrm{buffer}}}{Y_{\mathrm{buffer}}}
    + C_{\mathrm{package}},
  \label{eq:cost}
\end{equation}
where each die yield $Y$ follows the negative-binomial model
$Y = Y_{\mathrm{wafer}}\,(1 + A D_0/\alpha)^{-\alpha}$, and
$C_{\mathrm{package}}$ aggregates hybrid-bonding step cost and yield loss
across all stacking interfaces.  Defect densities are set to
$D_0 = 0.10\,\mathrm{cm}^{-2}$ (7\,nm) and
$0.05\,\mathrm{cm}^{-2}$ (45\,nm)~\cite{viksnewsletter2024}.
The host-facing PHY occupies $3.2\,\mathrm{mm}^2$ per channel and is
modeled as node-invariant.  Relocating this
$51\,\mathrm{mm}^2$ of interface silicon from the 7\,nm base die
(wafer cost \$9{,}346) to the 45\,nm buffer die (wafer cost \$2{,}000)
yields a $4.7\times$ reduction in per-unit wafer cost for the PHY alone.

Figure~\ref{fig:fig9} sweeps four parameters---hybrid-bonding cost per step, bonding yield, host-facing PHY area, and buffer-die area---to delineate the recurring-cost trade-off between Tri-Stack-PIM and an H$^2$-LLM-like two-die stacked design.

Overall, Tri-Stack-PIM remains cost-favorable provided per-step hybrid-bonding overhead stays below approximately \$5; beyond that threshold the additional bonding step erodes the saving (\ref{fig:fig9}a). Bonding yield, by contrast, exerts only marginal influence: across $Y_{\mathrm{bond}} \in [0.90,\,0.99]$ total cost shifts by less than 4\% (\ref{fig:fig9}b). The dominant lever is PHY I/O area (\ref{fig:fig9}c): each additional square millimeter costs $4.7\times$ more at 7\,nm than at 45\,nm, so the cost advantage widens monotonically with PHY size. Conversely, enlarging the buffer die raises $C_{\mathrm{buffer}}/Y_{\mathrm{buffer}}$ at 45\,nm and narrows the gap (\Cref{fig:fig9} (d)).

One cost factor lies outside this model. In a conventional two-die stack the base die is flip-chip bonded face-down to the interposer, presenting the host-facing PHY with a direct, well-characterized channel. Tri-Stack-PIM relocates the PHY to the buffer die, inserting an additional hybrid-bond interface and TSV tier into the host signal path. Accommodating this altered channel would necessitate PHY intellectual property (IP) re-design—an NRE and qualification expense that Eq.~(\ref{eq:cost}) does not capture. We leave quantification of this overhead to future work.

\noindent\textbf{Thermal Analysis.} High temperatures degrade DRAM retention, which raises the refresh rate and erodes effective bandwidth; reliable 3D-stacked design therefore requires the peak temperature to stay below \SI{85}{\degreeCelsius}~\cite{yue2024exploiting, yue20253d}. We perform a steady-state thermal simulation with HotSpot-7.0 in detailed\_3D grid mode~\cite{hotspot, hotspot3d}, modeling the full three-die stack (DRAM / buffer / base, each \SI{50}{\micro\meter} thinned Si) with a $64\times128$ spatial grid per layer. We construct a per-block floorplan and calculate each block's power at maximum utilization; the host-facing Peripheral Component Interconnect Express (PCIe) Gen5 PHY, the dominant heat source on the buffer die, dissipates up to \SI{3.58}{\watt} per stack (\SI{64}{GB/s} $\times$ \SI{7}{pJ/b}~\cite{bae2021pcie_gen4_tx, nvidia2022nvlink_c2c}). The total stack power under full load is \SI{11.97}{\watt}. We model the vertical heat resistance of the HB interface with an thermal conductivity of \SI{1.625}{\watt\per\meter\per\kelvin}~\cite{oprins20203d}. For the heat sink, we evaluate two schemes: advanced liquid cooling (\SI{0.05}{\kelvin\per\watt}) and microfluidic cooling (\SI{0.01}{\kelvin\per\watt})~\cite{stratum}. Under both, all three dies remain well below \SI{85}{\degreeCelsius}: Peak temperatures (DRAM / buffer / base) are \SI{63.74}{} / \SI{63.75}{} / \SI{62.63}{\degreeCelsius} with liquid cooling and \SI{61.90}{} / \SI{61.91}{} / \SI{60.76}{\degreeCelsius} with microfluidic cooling.

\begin{figure}[t]
  \centering
  \includegraphics[width=1\linewidth]{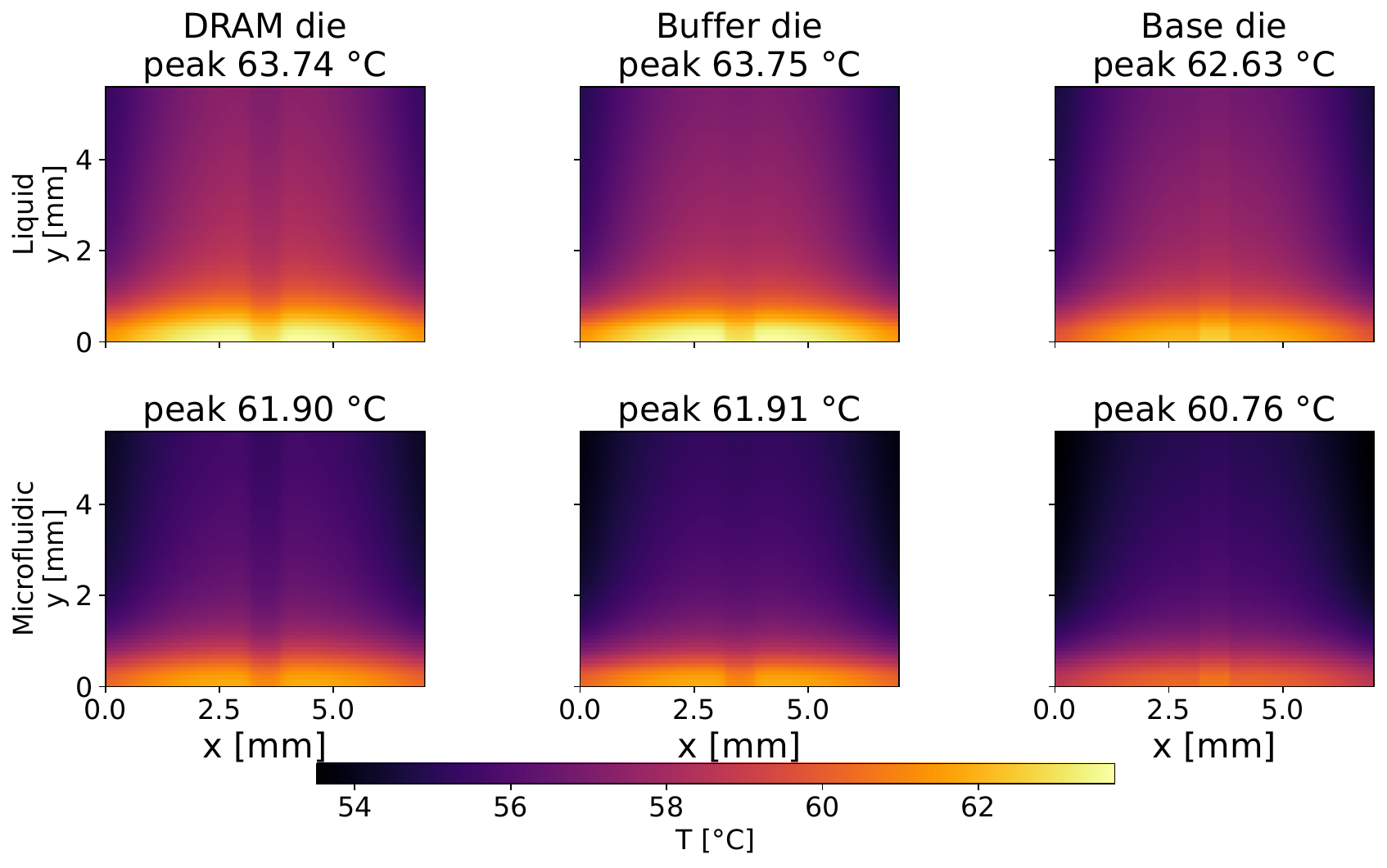}
  \caption{Steady-state temperature maps of the three-die stack.
    \textit{Top}: advanced liquid cooling.
    \textit{Bottom}: microfluidic cooling.
    Both remain \SI{22}{\kelvin} below the \SI{85}{\degreeCelsius} retention limit.}
\label{fig:thermal}
\end{figure}

\section{Conclusion}
\label{sec:conclusion}
We present Tri-Stack-PIM, a heterogeneous 3-die stacked architecture for LLM inference. Disaggregating DRAM, interface/control/SRE logic, and compute across three dies decouples interface scaling from compute provisioning. A dynamic scheduler on the Buffer Die reclaims head-of-line stall cycles via the UBAR priority policy, while a joint co-exploration yields balanced configurations and design insights. Tri-Stack-PIM achieves geometric-mean speedups of $4.95\times$, $2.57\times$, and $1.68\times$ over GPU-only, AttAcc, and H$^2$-LLM, respectively.

\begin{acks}
This work is supported by the National Natural Science Foundation of China
(No.~62404021), the Beijing Natural Science Foundation (No.~L2603021),
the Fundamental Research Funds for the BUPT (No.~2025AI4S20),
the Research Initiation Project for Introduced Talents of BUPT
(Nos.~510124058 and 510224062), and the Innovation Research Fund of BUPT.
\end{acks}

\makeatletter
\renewcommand{\bibsection}{%
  \vspace*{24pt}%
  \section*{\refname}%
  \phantomsection\addcontentsline{toc}{section}{\refname}%
}
\makeatother

{
    \bibliographystyle{ACM-Reference-Format}
    \balance
    \bibliography{ref/Top,ref/intro}
}

\end{document}